\documentclass[12pt]{article}
 
  \usepackage{amsfonts}
\usepackage[dvips]{graphicx}
\usepackage{latexsym}

\usepackage{amssymb,amsfonts,amsmath}
\usepackage{graphicx} 
\usepackage{indentfirst}

 \usepackage{bbm}

\newcommand {\cD}{{\cal D}}

\newcommand {\cF}{{\cal F}}

\newcommand {\cL}{{\cal L}}

\newcommand {\cN}{{\cal N}}

\newcommand {\cV}{{\cal V}}
\newcommand {\cW}{{\cal W}}

\def\a{\alpha}

\def\b{\beta}

\def\d{\delta}

\def\g{\gamma}
\def\G{\Gamma}

\def\l{\lambda}

\def\o{\omega}

\def\q{\theta}
\def\r{\rho}

\def\x{\xi}

\def\L{\Lambda}
\def\O{\Omega}

\def\S{\Sigma}

\def\X{\Xi}

\def\rd{{\rm d}}
\def\ri{{\rm i}}
\def\re{{\rm e}}

\newcommand{\ve}{\varepsilon}                            

\newcommand{\pa}{\partial}                           
\newcommand{\hf}{\frac12}

\newcommand{\vf}{\varphi}
\newcommand{\be}{\begin{equation}}
\newcommand{\ee}{\end{equation}}
\newcommand{\bea}{\begin{eqnarray}}
\newcommand{\eea}{\end{eqnarray}}

\newcommand{\1}{\underline{1}}
\newcommand{\2}{\underline{2}}

\def\double #1{#1{\hbox{\kern-2pt $#1$}}}

\newcommand{\hm}{{\hat{m}}}

\newcommand{\ha}{{\hat{a}}}
\newcommand{\hb}{{\hat{b}}}
\newcommand{\hc}{{\hat{c}}}

\newcommand{\hal}{{\hat{\a}}}
\newcommand{\hbe}{{\hat{\b}}}
\newcommand{\hga}{{\hat{\g}}}
\newcommand{\hde}{{\hat{\d}}}

\newcommand{\CN}{{\cal N}}

\newcommand{\bsm}{\begin{small}}
\newcommand{\esm}{\end{small}}

\begin{document}

\begin{center}
{\Large \bf  On 5D AdS SUSY and Harmonic Superspace}\footnote{Based on the talk 
given by G. T.-M.
at the Workshop "Supersymmetries and Quantum Symmetries" (SQS'07), Bogoliubov Laboratory of Theoretical Physics, JINR, Dubna, July 30-August 4 2007.}\\ 
\end{center}

\begin{center}

{\large  
Sergei M. Kuzenko
and 
Gabriele Tartaglino-Mazzucchelli\footnote{{\it E-mail addresses}: 
kuzenko@cyllene.uwa.edu.au (S. M. K.); gtm@cyllene.uwa.edu.au (G. T.-M.).}
} \\
\end{center}
\begin{center}
\footnotesize{
{\it School of Physics M013, The University of Western Australia\\
35 Stirling Highway, Crawley W.A. 6009, Australia}}  
~\\

\end{center}

\begin{abstract}
In our recent paper \cite{KT-M}, we presented the differential geometry 
of five-dimensional $\CN=1$ anti-de Sitter superspace 
AdS$^{5|8}$=SU(2,2$|$1)/SO(4,1)$\times$U(1)
and developed the harmonic and the projective superspace settings 
to formulate off-shell supersymmetric theories in AdS$^{5|8}$.
Here we give a brief review of 
the geometry and 
the harmonic superspace construction, and also elaborate on vector supermultiplets.
\end{abstract}


${}$For rigid supersymmetric theories with eight supercharges in $D\leq 6$ space-time dimensions, 
two powerful off-shell formalisms exist: 
harmonic superspace (HS) \cite{HarmonicSuperspace}
and projective superspace (PS) \cite{ProjectiveSuperspace}. 
HS is known to be useful, e.g., 
for the quantum computations in 4D $\CN=2, 4$ supersymmetric Yang-Mills theories. 
PS is  useful, e.g., for the explicit 
construction of  hyperk\"ahler metrics.
Unlike the case of rigid supersymmetry, there are still many open questions
as far as supergravity extensions of both approaches are concerned.  
In the HS setting, the prepotential structure of 4D $\CN=2$ supergravity is well understood 
\cite{SUGRA-har},  but a fully covariant formalism of differential geometry  
(similar to that existing for 4D $\cN=1$ supergravity) is still missing. 
As to a  PS approach  to supergravity-matter systems, 
it has only recently been developed in \cite{KT-M2} for five-dimensional $\cN=1$ supergravity, 
as a realization of the program put forward in \cite{KT-M}. 

The main thrust of \cite{KT-M} was to make use of the differential geometry of 
the 5D $\CN=1$ anti-de Sitter superspace AdS$^{5|8}$ (which is a maximally symmetric 
supergravity background) as the only technical input for developing 
HS and PS formalisms. In this note we briefly review the 
geometric properties of AdS$^{5|8}$ and  the HS construction of \cite{KT-M}. 

The 5D $\cN=1$ anti-de Sitter 
superspace AdS$^{5|8}$
can be identified with the coset space
SU(2,2$|$1)/SO(4,1)$\times$U(1).
The structure group of AdS$^{5|8}$ is SO(4,1)$\times$U(1), 
where SO(4,1) is the 5D Lorentz group. 
The superspace can be parametrized  by local coordinates 
 $z^{\hat{M}}=(x^{\hm},\q^{\hat{\mu}}_i)$,
 with $x^\hm$ bosonic coordinates, and $\q^{\hat{\mu}}_i$ fermionic coordinates 
 that form a 5D pseudo-Majorana spinor \bsm($\hm=0,..\,,4$, $\hat{\mu}=1,..\,,4$, $i=\1,\2$)\esm.  
 We also let a non-Abelian vector supermultiplet live in the superspace. 
 Then, the covariant derivatives can be represented in the form 
\bea
\cD_{\hat{A}}=E_{\hat{A}}~+~\hf \,\O_{\hat{A}}{}^{\hb\hc}\,M_{\hb\hc}
~+~{\rm i}\,\Phi_{\hat{A}}\,J~
+\ri\,\cal{V}_{\hat{A}}~.
\eea
Here $E_{\hat{A}}= E_{\hat{A}}{}^{\hat{M}}(z) \pa_{\hat{M}}$ is the supervielbein, 
with $\pa_{\hat{M}}={\pa/\pa z^{\hat{M}}}$;
 $M_{\hb\hc}$ and  $\O_{\hat{A}}{}^{\hb\hc}$ are the Lorentz  generators and 
 connection respectively (both antisymmetric in $\hb$, $\hc$); 
 $J$ and  $\Phi_{\hat{A}}$ are respectively the Hermitian U(1) generator and connection; 
 and $\cal{V}_{\hat{A}}$ is the Yang-Mills connection taking its values in the Lie algebra of the gauge group. 
The generators of the structure group act on the covariant derivatives as follows:
 \bea
{[}M_{\hal\hbe},\cD_{\hga}^i{]}\,=\,{1\over 2}\Big(\ve_{\hga\hal}\cD_{\hbe}^i 
+ \ve_{\hga\hbe}\cD_{\hal}^i\Big)~,\qquad 
{[}J,\cD_{\hal}^i{]}\,=\,J^i_{~j}\cD_{\hal}^j~,
\eea
where $M_{\hal\hbe}=M_{\hbe\hal}=(\S^{\ha\hb})_{\hal\hbe}M_{\ha\hb}$ 
(see the appendix in \cite{KT-M} for  our notation and conventions). 
The constant matrix $J^{ij}=\ve^{jk}J^i{}_k$ satisfies 
the conditions $J^{ij}= J^{ji},~\overline{J^{ij}} = -J_{ij}=-\ve_{ik}\ve_{jl}J^{kl}$. 
The covariant derivatives $\cD_{\hat{A}}$ transform under the Yang-Mills gauge group as 
$\cD_{\hat{A}} \to  \re^{\ri\tau}\,\cD_{\hat{A}}\,\re^{-\ri\tau}$, with $\tau(z)$
a Hermitian gauge  parameter.

The covariant derivatives obey the algebra 
($\cW=0$ in  the pure AdS$^{5|8}$ case  \cite{KT-M})
 \bea
\{\cD_{\hal}^i,\cD_{\hbe}^j\}=-2{\rm i}\ve^{ij}\cD_{\hal\hbe}
-4\o J^{ij}M_{\hal\hbe}-3\o\ve^{ij}\ve_{\hal\hbe}J
+2\ri\ve^{ij}\ve_{\hal\hbe}{\cal{W}}~,~~
\label{AlgCovDev1}~~~~~~\\
{[} \cD_{\ha}, \cD_{\hbe}^j {]}
={{\rm i}\over 2}\o J^j{}_{k}(\G_{\ha})_{\hbe}{}^{\hga}\cD^k_{\hga}
+\ri(\G_\ha)_\hbe{}^\hga\cD_\hga^j{\cal{W}}~,
~~
{[} \cD_{\ha},\cD_{\hb} {]}=-\o^2 J^2 M_{\ha\hb}+\ri{\cal{F}}_{\ha\hb}~,
\label{AlgCovDev2}\eea
with $\o$ a real constant, $J^2 \equiv \hf J^i{}_j J^j{}_i >0$ and 
$\cF_{\ha\hb}(z)={\ri\over 4}(\S_{\ha\hb})^{\hal\hbe}\cD^i_\hal\cD_{\hbe i}\cW$. 
The gauge field strength $\cW(z)$ satisfies the Bianchi identity
\be
\cD^{(i}_\hal\cD^{j)}_\hbe \cW={1\over 4}\ve_{\hal\hbe}\cD^{\hal(i}\cD^{j)}_\hal \cW~.
\label{BianchiID}
\ee
If the gauge group contains an Abelian factor, the corresponding vector 
multiplet can be frozen to possess a constant field strength, $\cW_0 ={\rm const}$, 
and this allows one to incorporate  a real central charge.

The isometry group of AdS$^{5|8}$ is SU(2,2$|1$). 
Its action on matter superfields is generated by Killing supervectors 
defined to be  those supervector fields
$ \x^{\hat{A}}(z) E_{\hat{A}}$ which enjoy the property 
\bea
\d_\x\cD_{\hat{A}}=-{[}(\x+\L^{\hbe\hga}M_{\hbe\hga}+{\rm i}\r J),\cD_{\hat{A}}{]} =0~,
\qquad 
\x \equiv\x^{\hat{A}}\cD_{\hat{A}}=\x^{\ha}\cD_{\ha}+\x^{\hal}_i\cD^i_{\hal}~.
\label{killingEq}
\eea
The parameters $\x^{\ha}(z),~\x^{\hal}_i(z),~\L^{\hbe\hga}(z)$ and $\rho(z)$ are constrained 
by eq. (\ref{killingEq}) 
in a nontrivial way \cite{KT-M}.
The isometry transformation of a covariant matter superfield $\chi$ 
on ${\rm AdS}^{5|8}$ 
is 
\bea
\d_\x\chi=-(\x+\L^{\hal\hbe}M_{\hal\hbe}+{\rm i} \,\r J)\chi~.
\label{fieldTransfGen}
\eea

In the HS approach  \cite{HarmonicSuperspace}, the superspace variables are extended 
to include SU(2) harmonics
$u^\pm_i$ 
subject to the constraints $(u^{+i})^*=u^-_i$ and $(u^+u^-)=1$ 
and defined modulo the equivalence relation $u^\pm \sim \exp (\pm {\rm i} \vf )\, u^\pm$.
In the AdS$^{5|8}$ case we choose the harmonics to be inert under the action of the operator $J$.
HS is then defined to be  AdS$^{5|8}\times S^2$ where the two-sphere is 
identified with the coset SU(2)/U(1). 

By using the harmonics $u^\pm_i$ we can switch to new bases for the isospinor indices 
and define objects like
$\cD^\pm_{\hal}\equiv u^{\pm}_i\cD^i_{\hal}$,
$J^{\pm\pm}\equiv u^\pm_i u^\pm_jJ^{ij}$.
In the new basis 
eq. (\ref{AlgCovDev1}) leads to
\bea
\{\cD_{\hal}^+,\cD_{\hbe}^+\}=-4\o J^{++}M_{\hal\hbe}~.
\label{twoD^+}
\eea
This  allows one  to define 
covariantly analytic superfields  
constrained by 
\bea
&&\cD^+_{\hal} Q^{(n)}(z,u)\,=\,0~,
\eea
provided  $Q^{(n)}$ is Lorentz scalar.
Such a superfield $Q^{(n)}$ is characterized  by 
an integer U(1) charge $n$, i.e. 
it  obeys the  condition
$Q^{(n)}(z,{\rm e}^{ \pm{\rm i} \vf } u^\pm)={\rm e}^{n {\rm i} \vf }Q^{(n)}(z,u^\pm)$.
In order to define a smooth tensor superfield over $S^2$, 
$Q^{(n)}$ should be given by a convergent  Fourier series  
\bea
Q^{(n)}(z,u)=\sum_{k=0}^{+\infty}Q^{(i_1\cdots i_{k+n}j_1\cdots j_k)}
(z)\,u^+_{i_1}\cdots u^+_{i_{k+n}}u^-_{j_1}\cdots u^-_{j_k}~.
\eea
The covariantly analytic superfield $Q^{(n)}$ 
transforms under the isometries of AdS$^{5|8}$ as
\be
\d_{\x}Q^{(n)}=-\Big(\x+{\rm i}\, \r J\Big) \,Q^{(n)}
=-\Big(\x^{\ha}\cD_{\ha}-\x^{+\hal}\cD^-_{\hal}+{\rm i}\,\r J\Big)Q^{(n)}~.
\ee
Here the operator  $J$ is realized by 
\bea
J \,Q^{(n)}(z,u)=\Big(J^{--}D^{++}
-J^{++}D^{--}+nJ^{+-}\Big) Q^{(n)}(z,u)~,
\eea
where $D^{++}=u^{+i} \,{\partial}/{\partial u^{- i}},~
D^{--}=u^{- i}\, {\partial}/{\partial u^{+ i}}$.
Given covariantly analytic superfields $Q^{(n)}$ and $Q^{(m)}$,
 their descendants 
 $Q^{(n+m)}:=Q^{(n)}Q^{(m)}$
and $Q^{(n+2)}:= D^{++}Q^{(n)}$
are also covariantly analytic. 

Let us give some examples of off-shell analytic multiplets living in AdS$^{5|8}$. 
The simplest one is an  $O(n)$ supermultiplet. It is  described by a superfield  
$H^{{i_1}\cdots{i_n}}(z)$ which is 
completely symmetric in its isospinor indices and 
satisfies the constraint $\cD^{(i_1}_\hal H^{{i_2}\cdots{i_{n+1}})}=0$. 
In the HS setting, associated with $H^{{i_1}\cdots{i_n}}(z)$ is 
$H^{(n)}(z,u)=u^+_{i_1}\cdots u^+_{i_n}H^{{i_1}\cdots{i_n}}(z)$, 
and the constraints on  $H^{{i_1}\cdots{i_n}}$ turn into
\be
\cD^+_\hal H^{(n)}=0~,\qquad D^{++}H^{(n)}=0~.
\ee
The $O(2)$ multiplet is used to describe an off-shell tensor multiplet, 
while $O(2n)$ multiplets with $n>1$ realize off-shell neutral hypermultiplets
with a finite number of auxiliary fields.

To describe an off-shell charged hypermultiplet in AdS$^{5|8}$, 
in complete analogy with the flat case \cite{HarmonicSuperspace}, 
one should make use of  a covariantly analytic superfield 
$q^+$, $\cD^+_\hal q^+=0$. 

As another important  example,  let us consider an Abelian vector multiplet. 
Its HS description is provided by a real covariantly analytic prepotential 
$V^{++}$, $\cD^+_\hal V^{++}=0$, 
defined modulo  Abelian gauge transformations of the form
\be
\d V^{++}=-D^{++}\l~,
\label{gaugeTransf}
\ee
with the gauge parameter $\l$ being covariantly analytic, $\cD^+_\hal\l=0$. 
Associated with  $V^{++}$
is the field strength $W(z)$ 
(compare with the flat case \cite{5DflatHS})
\be
W(z)={\ri\over 8}\int \rd u\,(\hat{\cD}^-)^2\,V^{++}(z,u)~,
\label{gaugeFS}
\ee
where $(\hat{\cD}^\pm)^2\equiv{\cD}^{\hal\pm}{\cD}^\pm_{\hal}$. 
Using the fact that $\int\rd u \,J Q^{(0)}=0$ is zero for any superfield of charge zero, 
it is simple to prove that $W$ is invariant under the gauge transformations 
(\ref{gaugeTransf}) and that it transforms under the isometry group SU(2,2$|$1) as 
$\d_\x W=-\x W$. Furthermore, it can be seen that $W$ satisfies the Bianchi identity (\ref{BianchiID}).

Having defined off-shell supermultiplets living in AdS$^{5|8}\times S^2$,
the next issue is to derive a supersymmetric action principle
in harmonic superspace.  It turns out that, 
for any  covariantly analytic Lagrangian $\cL^{(+4)}$ of charge $+4$,
$\cD^+_{\hal}\cL^{(+4)}=0$,
the action \cite{KT-M}
\bea
S \big(\cL^{(+4)} \big)=
\int \rd^5x\,e \int \rd u\,\Big{[}(\hat{\cD}^-)^4+{2\over 3}\o J^{--}(\hat{\cD}^-)^2\Big{]}\cL^{(+4)}\Big|~,
\qquad e\equiv\det(e_\hm{}^\ha)~,
\label{SUSYactionHS}
\eea
is invariant under the isometry group SU$(2,2|1)$, 
$\d_\x S\equiv 0$.  Here the operator measure involves 
the fourth-order operator $(\hat{\cD}^-)^4=-{1\over 96}\ve^{\hal\hbe\hga\hde}
\cD^-_{\hal}\cD^-_{\hbe}\cD^-_{\hga}\cD^-_{\hde}$.
The action (\ref{SUSYactionHS}) is evaluated in the Wess-Zumino gauge \cite{KT-M}
in which 
\be
\cD_{\hat{a}} | = \nabla_{\hat a} = e_{\hat a}{}^{\hat m} (x) \, \pa_{\hat m} 
+ \hf \o_{\hat a}{}^{\hat b \hat c} (x) \,M_{\hat  b \hat c}+\ri V_\ha (x)~,~~~~~~
{[} \nabla_{\ha},\nabla_{\hb} {]}=-\o^2 J^2 M_{\ha\hb}+\ri F_{\ha\hb}~,
\ee
where $\nabla_{\hat a}$ are the covariant derivatives of 5D anti-de Sitter space 
in the presence of the Yang-Mills
connection $V_\ha=\cV_\ha|$, 
with $F_{\ha\hb}=\cF_{\ha\hb}|$ the corresponding field strength.

The above action was found  in \cite{KT-M} by starting from a general  
ansatz for the operator measure of the form
$[(\hat{\cD}^-)^4+a_1\o J^{--}(\hat{\cD}^-)^2+a_2(\o J^{--})^2]$,  
with $a_{1,2}$ some $\o$-independent  coefficients, 
and then 
imposing the condition  $\d_\x S\equiv 0$, for an arbitrary AdS Killing supervector $\x$.
The latter condition led us to the unique  choice $a_1=2/3$ and $a_2=0$.

The important property of our action (\ref{SUSYactionHS}), 
which was not discussed in \cite{KT-M},  is  that $S\big(D^{++}\X^{++}\big)=0$ for any 
covariantly analytic superfield $\X^{++}$.
Unlike the flat superspace case,
this property is rather nontrivial in  AdS$^{5|8}\times S^2$.
It turns out  that each of the conditions $\d_\x S=0$ and $S\big(D^{++}\X^{++}\big)=0$
holds if and only if the coefficients $a_{1,2}$ in the operator measure 
are chosen to be $a_1=2/3$ and $a_2=0$.
In this sense, the property $S\big(D^{++}\X^{++}\big)=0$ is similar to the principle of projective 
invariance advocated in \cite{KT-M,KT-M2}.

The action principle (\ref{SUSYactionHS}) allows one to generalize
many flat superspace models to the AdS$^{5|8}$ case, and here we present  an example of
Chern-Simons couplings for a set of Abelian vector multiplets $V^{++}_I$, where  $I=1,\cdots,n$.
Due to (\ref{BianchiID}),
the corresponding field strengths $W_I$ obey
$\cD^+_\hal\cD^+_\hbe \cD^+_\hga W_I=2\o\ve_{\hbe\hga}J^{++}\cD^+_\hal W_I$.
This implies that 
\be
G^{++}_{IJ}=G^{++}_{JI}
=\ri\Big{\{}\cD^{+\hal}W_I\cD^+_\hal W_J+\hf W_{(I}(\hat{\cD}^+)^2W_{J)}-2\o J^{++}W_IW_J\Big{\}}~,
\ee
enjoys the properties  $\cD^+_{\hat \a}G^{++}_{IJ}=D^{++}G^{++}_{IJ}=0$ and, hence, is  an $O(2)$ multiplet. 
Then, a supersymmetric action invariant under gauge transformations 
$\d V^{++}_I=-D^{++}\l_I$ is generated by the Lagrangian
\be
\cL^{(+4)}={1\over 12}c_{I,JK}V^{++}_IG^{++}_{JK}~,
\ee
with some coupling constants $c_{I,JK}$ (compare with the flat superspace case \cite{5DflatHS}).
\\

\noindent
{\bf Acknowledgements:}
This work is supported in part by the Australian Research Council and by a UWA research grant.


\small{

}

\end{document}